\documentclass[12pt,fleqn]{article}
\usepackage{amsfonts,times,epsfig}

\newcommand{\dA}{{\dot{A}}}
\newcommand{\dB}{{\dot{B}}}

\newcommand{\hQ}{{\hat{Q}}}
\newcommand{\hP}{{\hat{P}}}
\newcommand{\hD}{{\hat D}}

\newcommand{\CO}{{\cal{O}}}

\newcommand{\ft}[2]{{\textstyle {\frac{#1}{#2}} }}

\newcommand{\tr}{{\rm tr \,}}

\newcommand{\ra}{\rightarrow}

\newcommand{\e}{\varepsilon}

\def\C{\mathbb{C}}

\def\R{\mathbb{R}}
\def\N{\mathbb{N}}

\newcommand{\be}{\begin{equation}}
\newcommand{\ee}{\end{equation}}
\newcommand{\ben}{\begin{displaymath}}
\newcommand{\een}{\end{displaymath}}
\newcommand{\ba}{\begin{eqnarray}}
\newcommand{\ea}{\end{eqnarray}}
\newcommand{\nn}{\nonumber}
\newcommand{\non}{\nonumber\\}
\newcommand{\bean}{\begin{eqnarray*}}
\newcommand{\eean}{\end{eqnarray*}}

\newcommand{\mathon}{\mathversion{bold}}
\newcommand{\mathoff}{\mathversion{normal}}

\newcommand{\Ref}[1]{(\ref{#1})}

\makeatletter
\@addtoreset{equation}{section}
\makeatother

\newcommand{\pls}{\!+\!}
\newcommand{\mis}{\!-\!}

\newcommand{\eee}{\mathfrak{e}_{8(8)}}
\newcommand{\eneun}{\mathfrak{e}_{9}}
\newcommand{\ke}{\mathfrak{ke}_{9}}

\newcommand{\EE}{{\rm E}_{8(8)}}
\newcommand{\Eneun}{{\rm E}_{9(9)}}
\newcommand{\KE}{{K({\rm E}_9)}}
\newcommand{\sr}{\gamma}

\newcommand{\CC}{{\cal C}}
\newcommand{\CX}{{\cal X}}
\newcommand{\CE}{{\cal E}}
\newcommand{\hCV}{\hat{\cal{V}}}
\newcommand{\CV}{{\cal{V}}}
\newcommand{\CM}{{\cal{M}}}
\newcommand{\CD}{{\cal D}}
\newcommand{\CP}{{\cal{P}}}
\newcommand{\CQ}{{\cal Q}}
\newcommand{\CJ}{{\cal J}}

\newcommand{\cI}{{\cal I}}
\newcommand{\cO}{{\cal O}}
\newcommand{\hL}{\hat{\Lambda}}
\newcommand{\trho}{\tilde{\rho}}

\newcommand{\I}{{\rm i}}

\begin{document}

\thispagestyle{empty}

\begin{flushright}
AEI-2004-029\\
DESY 04-119
\end{flushright}
\bigskip

\begin{center}

\mathon
{\bf\Large On $K({\rm E}_9)$}
\medskip
\mathoff

\vspace{10mm}

{\bf H. Nicolai}

{{\em Max-Planck-Institut f{\"u}r Gravitationsphysik,
Albert-Einstein-Institut,\\
 M\"uhlenberg 1, D-14476 Potsdam, Germany}}\\
{\small Hermann.Nicolai@aei.mpg.de} 

\vspace{.5cm}
{\bf H.~Samtleben}

{{\em II. Institut f\"ur Theoretische Physik der Universit\"at Hamburg,\\ 
 Luruper Chausse 179, D-22761 Hamburg, Germany}}\\
{\small Henning.Samtleben@desy.de} 
\end{center}

\medskip
\begin{abstract}
We study the maximal compact subgroup $\KE$ of the affine Lie 
group $\Eneun$ and its on-shell realization as an R symmetry of 
maximal $N=16$ supergravity in two dimensions. We first give a 
rigorous definition of the group $\KE$, which lives on the double 
cover of the spectral parameter plane, and show that the infinitesimal 
action of $\KE$ on the chiral components of the bosons and the fermions 
is determined in terms of an expansion of the Lie algebra of $\KE$ 
about the two branch points of this cover; this implies in particular 
that the fermions of $N=16$ supergravity transform in a spinor 
representation of $\KE$. The fermionic equations of motion can be fitted 
into the lowest components of a single $\KE$ covariant `Dirac equation', 
with the linear system of ${N=16}$ supergravity as the gauge connection. 
These results suggest the existence of an `off-shell' realization 
of $\KE$ in terms of an infinite component spinor representation. 
We conclude with some coments on `generalized holonomies' 
of M theory.

\end{abstract}

\renewcommand{\thefootnote}{\arabic{footnote}}
\setcounter{footnote}{0}
\renewcommand{\thefootnote}{\arabic{footnote}}

\newpage

\section{Introduction}

The R symmetries of maximally extended supergravities are the maximal
compact subgroups of the global ${\rm E}_{n(n)}$ symmetries known or
conjectured to arise in the dimensional reduction of $D=11$
supergravity with $n$ (spacelike) Killing vectors to $11\!-\!n$
dimensions~\cite{CreJul79,Juli80}.  For all $n$, these are defined as
the invariant subgroups w.r.t.~the Chevalley involution $\theta$,
which is defined by its action on the Chevalley generators (see
e.g.~\cite{Kac})
\be\label{Inv}
\theta (e_i) = - f_i \;\; , \quad
\theta (f_i) = - e_i \;\; , \quad
\theta (h_i) = - h_i 
\;,
\ee
In three or more space-time dimensions, $n\leq 8$, and these groups,
which we will denote here generally by $K({\rm E}_n)$, are finite
dimensional and well understood. By contrast, in less than three
dimensions, where $n\geq 9$, ${\rm E}_{n(n)}$ and $K({\rm E}_n)$ are
both infinite dimensional groups. In this paper, we will focus on the
case $n=9$, i.e.~the Lie group $\KE\subset\Eneun$ and its associated
involutory Lie algebra $\ke:= {\rm Lie} \, \big(K({\rm E}_9)\big)
\subset \eneun:= {\rm Lie} \, ({\rm E}_{9(9)}) $\, \footnote{The group 
$\KE$ is sometimes also designated by ${\rm SO}(16)^\infty$.}; more 
specifically, we will study the realization of this symmetry in the 
context of maximal $N=16$ supergravity in two dimensions. This case
is still far simpler than $n \geq 10$, for which the Cartan matrices
of ${\rm E}_n$ become indefinite. However, even though the affine Lie group 
${\rm E}_{9(9)}$ and its realization as a loop group are fairly well
understood, very little seems to be known about its compact 
subgroup $\KE$ or the irreducible representations of $\KE$: 
for instance, the standard textbook on loop groups \cite{PreSeg86} 
contains no information on this topic. Involutory subalgebras of
Kac Moody algebras and their relation with Slodowy algebras were 
studied in \cite{Berm89}\footnote{We are grateful to A. Kleinschmidt 
for bringing this reference to our attention.}, but these require
in addition outer automorphisms of the Dynkin diagram, which the
${\rm E}_9$ algebra does not possess. Furthermore, by all appearances,
the subject of {\em spinorial (i.e.~double valued) representations}   
of $\KE$ (or any other infinite dimensional involutory subgroup of 
an affine Lie group) is mostly {\em terra incognita}. 

Our study is motivated by the structure of the linear systems 
for gravity and supergravity in two dimensions, and by the desire
to understand the as yet unknown infinite dimensional symmetries 
underlying string and M theory. For the bosonic theories, it has been 
known for a long time from the study of the Geroch group in general 
relativity and its generalizations, that the affine symmetries arising
in the reduction to two dimensions can be realized on an infinite set 
of `dual potentials' which in the non-linear realization are non-linear 
and non-local functions of a finite number of physical fields (see 
\cite{DieHoe84} for a review from the relativist's perspective, 
and \cite{BreMai87} for a more general treatment emphasizing the 
group theoretical aspects). 
These dual potentials are known to arise from the solution of the 
associated linear system via an expansion in the spectral parameter 
$\sr$ about the point $\sr=0$. For the locally supersymmetric models 
and their linear systems, which were investigated in 
\cite{Nico87a,NicWar89,NicSam98}, a direct fermionic analog of the 
bosonic dual potentials does not appear to exist, for the very simple
reason that fermions obey first order equations of motion, and hence
cannot be dualized. In this paper, we will present evidence that an
infinite hierarchy of fermionic fields nevertheless does exist.

In more mathematical terms, we propose that the fermionic fields
belong to a {\em spinor representation of} $\KE$, which arises not by
dualization or an expansion about $\sr =0$, but by an expansion about
the two branch points $\sr=\pm 1$ in the spectral parameter plane,
which are associated to the positive and negative chirality components
of the fermionic fields. Our main point here is the observation that
the fermionic multiplet considered in \cite{NicSam98}, which contains
the fermion fields of $N=16,d=2$ supergravity, already by itself
constitutes a spinorial, albeit non-faithful, representation of
$\KE$. In this paper, we will work out these transformations in more
detail, reducing the complicated contour integrals of \cite{NicSam98}
to the much simpler formulas \Ref{E9ff}, \Ref{E9ff1}. As we will
furthermore show in section~6, the fermionic equations of motion can
be combined into a single $\KE$ covariant 'Dirac equation', with the
linear system of $N=16$ supergravity~\cite{Nico87a,NicWar89,NicSam98}
serving as the $\KE$ connection.

A crucial distinction that we will make in this paper is between
`on-shell' and `off shell' realizations of $\KE$. By the former we
mean those (field and coordinate dependent) $\KE$ transformations
leaving the $N=16$ supergravity equations of motion form invariant.
By an `off shell' realization, on the other hand, we mean a
realization in terms of infinitely many fields, that does not require
the equations of motion to be satisfied. In section~4 we will explain
how such a representation can be arrived at for the bosonic theory.
We will describe the peculiar difficulties that one encounters when
trying to construct `off shell' spinor representations of $\KE$ in
section~6. Although we are not able so far to give a complete
characterization of the latter, we believe that the present results
constitute a first step towards their systematic construction.

\section{Preliminaries}

For the reader's convenience, we start out with a summary of some 
pertinent facts about the linear systems and spectral parameters
appearing in $2d$ gravity and supergravity, following~\cite{BreMai87}
and \cite{Nico91}, before giving a rigorous definition of the 
groups $\Eneun$ and $\KE$ in the next section.

\subsection{Spectral parameter}
We briefly recall some basic facts about the spectral parameters 
appearing in the linear systems for $2d$ gravity and supergravity, 
see~\cite{BreMai87,Nico91,NicSam98} for our conventions and more 
details. For Lorentzian worldsheets, the spectral parameter~$\sr$ 
entering the linear system is introduced as the function 
\be\label{sr}
\sr(\rho,\trho;w) = \frac1{\rho}\left(w+\trho-\sqrt{(w+\trho)^2-\rho^2}
\;\right)\; ;
\ee
it depends on the $2d$ coordinates via the `dilaton' $\rho$ and 
the `axion' $\trho$, which are dual to one another:
\footnote{With $x^\pm := \frac1{\sqrt{2}}(x^0 \pm x^1)$ and
$\partial_\pm := \frac1{\sqrt{2}}(\partial_0 \pm \partial_1)$.}
\be\label{boxrho}
\partial_\pm \rho = \pm \partial_\pm \trho
\;,
\ee
and hence both obey free field equations of motion
$\partial_+ \partial_- \rho = \partial_+ \partial_- \trho = 0$.

The third variable $w$ arises as a constant of integration in the 
first order differential equations satisfied by $\sr(\rho,\trho;w)$ 
\be\label{spdiff1}
\sr^{-1}\partial_\pm\sr = \frac{1\mp\sr}{1\pm\sr}\,\rho^{-1}
\partial_\pm\rho\;,
\ee
which are compatible by virtue of (\ref{boxrho}). We will occasionally
refer to $\sr$ and $w$ as the `variable' and the `constant' 
spectral parameter, respectively, as both play an important role 
in understanding the symmetries of $2d$ (super)gravity. From \Ref{sr}, 
we immediately obtain
\be\label{gdg}
\sr^{-1}\partial_w \sr ~=~ -\frac{2\sr}{\rho(1-\sr^2)} \;,
\ee
and the inverse relation
\ba\label{w}
w(\sr)&=&\frac{\rho}2 \left(\sr\!+\!\frac1{\sr}\right)-\trho
\;.
\ea

{}From \Ref{sr} we see that the function $\sr (w)$ lives on a 
two-sheeted cover of the complex $w$-plane, with the branch points
\be
\sr = \pm 1  \; \Longleftrightarrow \; 
w = \pm \rho - \trho
\;.
\ee
Notice that these are `moving' branch points, as they depend on the
spacetime coordinates via the fields $\rho$ and $\trho$. At the same 
time they are fixed points of the involution
\be\label{Involution}
\cI : \sr \longrightarrow \frac{1}{\sr} \qquad
\Longrightarrow \qquad w = w \circ \cI
\;,
\ee 
which exchanges the two sheets. The branch points play a special role 
because we will show that they are associated with the two chiralities of 
the fermions and the bosonic currents on the world sheet. For later use, 
we therefore introduce the new variables
\ba
u_\pm = u_\mp^{-1} =
\frac{1\!\pm\!\gamma}{1\!\mp\!\gamma} =
\sqrt{\frac{w + \trho \pm \rho}{w + \trho \mp \rho}}
\qquad \Longrightarrow \qquad
\gamma = \mp\frac{1\mis u_\pm}{1\pls u_\pm}\;,
\ea
which may be viewed as local coordinates in the $\sr$ plane around the 
two branch points $u_+ =0$ and $u_- =0$, respectively (with $u_\pm = \infty$ 
at the opposite branch point). With this Eq.~\Ref{spdiff1} becomes
\be
\sr^{-1} \partial_\pm \sr = u_\pm^{-1} \rho^{-1}\partial_\pm \rho
\;.
\ee
Let us record the following formulas which will be useful later
\be\label{udu}
\gamma\,\partial_\gamma = \ft12(u_\pm^2\mis1)\,
\partial_{u_\pm} \;, \qquad
u_\pm^{-1} \partial_\pm u_\pm = - u_\mp^{-1} \partial_\pm u_\mp =
\ft12 (1- u_\pm^{-2}) \rho^{-1}\partial_\pm \rho
\;,
\ee
and
\be
v-w ~=~ \frac{\rho}{2}\:\frac{\big(\sr(v)\!-\!\sr(w)\big)\,
\big(\sr(v)\sr(w)\!-\!1\big)}
{\sr(v)\sr(w)}\;.
\ee

\subsection{Linear system}

As explained in \cite{BreMai87,Nico87a,NicWar89,NicSam98}, the linear 
system of $N=16$ supergravity in two spacetime dimensions is formulated 
in terms of an ${\rm E}_8$ matrix $\hCV(x,\sr)$ which depends on the 
spacetime coordinates $x\equiv (x^0,x^1)$ (or $x\equiv (x^+, x^-)$) and 
the variable spectral parameter $\sr$, and which for real values of $\sr$ 
is an element of the real form $\EE$. It is subject to the transformations
\be\label{E9KE9} 
\hCV(x,\sr)\longrightarrow G(w) \hCV(x,\sr) H(x,\sr) 
\;,
\ee
where $G(w)$ is an element of the loop group ${\rm E}_{9(9)}$, and 
$H(x,\sr)$ an element of its `maximal compact' subgroup $K({\rm E}_9)$.
We will properly define these groups in the following section,
but here already note that the latter consists of those ${\rm E}_{8(8)}$ 
valued functions $H(\gamma)$ satisfying
\be
\label{KE9}
H(\gamma)^{-1} =  H(1/\gamma)^T \;.
\ee
The matrix $\hCV$ satisfies the linear partial differential equations
\be\label{ls}
\partial_\pm \hCV(x,\sr) = \hCV(x,\sr) L_\pm(x,\sr) 
\;,
\ee
with the Lax connection $L_\pm(x,t,\sr)$ defined as function of
the physical fields. In the absence of fermionic fields, 
the Lax connection is \cite{BreMai87}
\be
L_\pm(x,\sr) = Q_\pm (x)+ \frac{1\mp\sr}{1\pm\sr}\,P_\pm (x) \in\eee\;.
\label{Lbos}
\ee
The bosonic currents $Q_\pm\in \mathfrak{so}(16)$ and 
$P_\pm \in \eee$ are defined by
\ba\label{VdV}
\CV^{-1} \partial_\pm \CV &\equiv& Q_\pm + P_\pm 
~\equiv~   \ft12\, Q_\pm^{IJ}\,X^{IJ} + P_\pm^A\,Y^A \;,
\ea
where $\CV$ is an ${\rm E}_{8(8)}$ valued matrix in which the
bosonic fields of the theory are assembled. The $X^{IJ}$ and $Y^{A}$
denote the 120 compact and 128 noncompact generators of $\eee$, 
respectively (see~\cite{NicSam98} for more detailed explanation
of our notations and conventions). The equations \Ref{ls} are
compatible if and only if the equations of motion of the original 
${\rm E}_{8(8)}/{\rm SO}(16)$ $\sigma$-model are satisfied.
In addition to these second order equations of motion, the model
exhibits the first order {\em conformal constraints}
\ba 
 T_{\pm\pm}\equiv\ft12\rho\, P_\pm^A P_\pm^A 
- \partial_\pm\rho\,\partial_\pm\sigma ~\approx~ 0
\;,
\label{TTT}
\ea
that determine the conformal factor $\sigma$ of the two-dimensional
metric in terms of the matter currents.

The fermionic fields of $N=16$ supergravity are the nonpropagating gravitino 
fields $\psi_\mu^I$, the dilatino fields $\psi_2^I$, which both originate
from the gravitino in three dimensions, and the spin-$1/2$ fields 
$\chi^\dA$, transforming in the ${\bf 16}$ and ${\bf 128_c}$ of 
${Spin}(16)$, respectively. With the chiral notation of \cite{NicSam98} 
the full Lax connection is given by
\ba\label{LS1}
L_\pm(\sr)&\equiv&
\ft12\hQ^{IJ}_\pm(\sr)X^{IJ}+\hP^{A}_\pm(\sr)Y^A \;,
\non
\hQ^{IJ}_\pm(\sr) &=& 
Q^{IJ}_\pm -\frac{2\I\sr}{(1\!\pm\!\sr)^2}\; 
\left(8\psi_{2\,\pm}^{[I}\psi^{J]}_\pm \pm \Gamma^{IJ}_{\dot{A}\dot{B}}
\chi_\pm^{\dot{A}} \chi_\pm^{\dot{B}} \right) 
-\frac{32\I\sr^2}{(1\!\pm\!\sr)^4}\;
\psi_{2\,\pm}^{I}\psi_{2\,\pm}^{J} \;,\non
\hP^{A}_\pm(\sr) &=& \frac{1\!\mp\!\sr}{1\!\pm\!\sr}\;P_\pm^A
+\frac{4\I\sr(1\!\mp\!\sr)}{(1\!\pm\!\sr)^3}\;
\Gamma^{I}_{A\dot{B}}\psi_{2\,\pm}^{I}\chi^{\dot{B}}_\pm 
\;,
\ea
with ${\rm SO}(16)$ $\Gamma$-matrices $\Gamma^I_{A\dA}$. The 
compatibility equations of \Ref{ls} with this connection reproduce the
full supergravity equations of motion including fermionic terms to all 
orders~\cite{NicSam98}. For later use, let us also write out the 
linear system in terms of the coordinates $u_\pm$:
\ba\label{LS3}
\hQ^{IJ}_\pm(u_\pm) &=& Q^{IJ}_\pm  \pm \frac{\I}2 \left(u_\pm^{-2} -1\right) 
\left(8\psi_{2\,\pm}^{[I}\psi^{J]}_\pm \pm \Gamma^{IJ}_{\dot{A}\dot{B}}
\chi_\pm^{\dot{A}} \chi_\pm^{\dot{B}} \right) 
- 2\I \left(u_\pm^{-2} -1\right)^2 
\psi_{2\,\pm}^{I}\psi_{2\,\pm}^{J} \;,\non
\hP^{A}_\pm(u_\pm) &=& u_\pm^{-1}\;P_\pm^A 
\mp  \I u_\pm^{-1} \left(u_\pm^{-2} -1 \right)
\Gamma^{I}_{A\dot{B}}\psi_{2\,\pm}^{I}\chi^{\dot{B}}_\pm 
\;.
\ea
The linear system thus is singular at the branch points $u_\pm =0$,
with the bosonic chiral currents appearing as the coefficients
of the first order pole, and the fermionic bilinears as the
coefficients of the higher order poles. By local supersymmetry
we can set the dilatino $\psi_2^I=0$, in which case the linear
system has only a second order pole multiplying a fermionic 
bilinear in the fields $\chi^\dA_\pm$ in addition to the first
oder pole from the bosons \cite{Nico87a}. The conformal constraints 
\Ref{TTT} receive quadratic and quartic corrections in the fermions.
In addition there are the {\em superconformal constraints}
\ba
{\cal S}^I_\pm &\equiv& D_\pm(\rho\psi_{2\,\pm}^I) 
-\rho\partial_\pm\sigma\psi_{2\,\pm}^I 
\mp
\rho\,\Gamma^I_{A\dA} \,\chi^\dA_\pm\,P^A_\pm
\pm\partial_\pm\rho\,\psi_\pm^I
~\approx~0
\;,
\label{TS}
\ea
modulo cubic fermion terms\footnote{The full expression for the superconformal
constraints including all higher order fermionic terms
has been worked out in~\cite{NicSam98}.}, 
that allow to express the gravition field $\psi_\pm^I$ in terms of 
the other fields. 

For the bosonic theory, it was shown in
\cite{BreMai87} that the phase space of the two-dimensional theory 
--- in this case, the space of solutions of the equations of motion 
of $D\!=\!11$ supergravity with nine commuting Killing vectors ---
can be described in terms of the infinite-dimensional coset space
\ba
\label{GH}
{\rm G}/{\rm H} &=& {\rm E}_{9(9)}/K({\rm E}_9)
\;,
\ea
parametrized by the matrices $\hCV(x,\sr)$. In accordance with
\Ref{E9KE9}, the global $\Eneun$ symmetry acts from the left on this
coset space, while the local $\KE$ symmetry acts on it from the right.
The latter may be used to bring $\hCV$ into a generalized 
triangular gauge, defined by requiring $\hCV$ to be holomorphic
in a neighborhood of $\sr =0$ \cite{Juli83,BreMai87}. This gauge
choice allows to read off the solution of the bosonic field
equations by setting $\sr =0$:
\be
\CV(x) = \hCV(x,\sr) \Big|_{\sr =0}
\;.
\ee
In the triangular gauge, $\hCV$ can be represented in the form
\ba
\hCV(x,\gamma) &=& S(w)H(x,\gamma) \;,
\ea
where $\hCV(\gamma)$ is holomorphic inside the unit disc in the complex 
$\gamma$-plane, and $H(x,\sr)$ belongs to $\KE$, i.e.~obeys \Ref{KE9}.
This fixes $H(x,\gamma)$ up to constant ${\rm SO}(16)$ transformations.
On such $\hCV$, the global non-linear and non-local action of $\Eneun$ 
takes the (infinitesimal) form
\be\label{actionVBM}
\delta_\Lambda\,\hCV(x,\sr(w)) ~=~
\Lambda(w)\hCV(x,\sr(w))-
\hCV(x,\sr(w))\,\Upsilon_{\!\Lambda}(x,\sr(w))\;,
\ee
Here, $\Lambda(w)$ is an $\mathfrak{e}_8$ valued, but coordinate 
independent, function on the complex $w$-plane parametrizing the 
infinitesimal action of $\mathfrak{e}_9\equiv {\rm Lie}({E}_9)$. 
The matrix $\Upsilon_{\!\Lambda}(x,\sr(w))$ is a special field
dependent element of the algebra $\mathfrak{ke}_{9}\equiv 
{\rm Lie}(K({\rm E}_9))$: it restores the holomorphic gauge which 
is violated by the pure action of~$\Lambda(w)$ in~\Ref{actionVBM}. 
On the physical fields, $\Eneun$ thus acts by non-linear and non-local 
transformations. The loop algebra is recovered by chosing 
$\Lambda(w)=\Lambda_n w^n$. It follows from \Ref{w} that compensating 
$\KE$ transformations are only required for $n\geq 0$, while transformations 
with $n<0$ do not violate the triangular gauge; the latter only shift 
the dual potentials by constants (and are thus related to the integration
constants arising at each step of the dualization), and have no
effect on the physical fields.

\mathon
\section{$\Eneun$ and $\KE$}
\mathoff

We now give a rigorous definition of the two groups $\Eneun$ and 
$\KE$, following \cite{BreMai87,JulNic96}. In view of possible later
applications, we will aim for a `coordinate free' description, 
where we can regard $w$ and $\sr$ merely as local coordinates 
on some Riemann surface $\Sigma$, and its double cover $\tilde\Sigma$, 
respectively (although this is not strictly necessary for the 
purposes of the present paper). The involution $\cI$ exchanges
the two sheets of the cover; on $\tilde\Sigma$, it generalizes 
\Ref{Involution} in such a way that its two fixed points map to 
the branch points of $\tilde\Sigma$. For stationary axisymmetric 
or colliding plane wave solutions, $\Sigma$ is just the complex 
plane, but there do exist solutions for which $\Sigma$ is a 
hyperelliptic Riemann surface, namely the algebro-geometric 
solutions of \cite{Koro89,KorMat90}. One can discuss the Ernst equation 
and its generalizations on even more general Riemann surfaces \cite{KorNic94},
although the spectral problem is not completely understood in
that case. Nevertheless, we will keep the discussion in this section 
quite general with these possible generalizations in mind.

Viewing $w$ and $\sr$ as local coordinates, it was shown in~\cite{JulNic96}
how to reformulate the defining equation \Ref{sr} as a linear system 
taking values in a certain subalgebra of the Lie algebra of vector 
fields on $\tilde\Sigma$. For this purpose, one generalizes the inverse
relation \Ref{w} by considering coordinate dependent maps
\be
Y: \tilde\Sigma \longrightarrow \Sigma \qquad \mbox{with} \quad
Y\circ \cI = Y
\;.
\ee
The variable $Y$ here corresponds to the variable $w^{-1}$, and the 
maps $Y$ should be thought of as generalizing the relation \Ref{w}. 
The $w$-diffeomorphisms $f$ and the $\sr$-diffeomorphisms $k$ then act 
from left and right on $Y$ according to
\be
Y \longrightarrow f\circ Y \circ k \qquad \mbox{for} \quad
f\in{\rm Diff}^+ (\Sigma) \quad , \quad k \in {\rm Diff}^+ (\tilde\Sigma)
\;.
\ee
In order to preserve the double cover, we demand that the diffeomorphims 
$k$ acting on $\sr$ satisfy
\be\label{kdiff}
k\circ \cI = \cI \circ k \; ,
\ee
ensuring in particular that $k$ preserves the two branch points. 
The associated Lie algebra is the involutory subalgebra of the 
Witt algebra generated by \cite{JulNic96}
\be\label{Kn}
{\cal K}_n = \left( - \sr^{n+1} + \sr^{-n+1}\right) {\partial}_{\sr} 
\;.
\ee
These generators define a maximal `anomaly free' subalgebra (i.e.~without 
central extension) of the Witt-Virasoro algebra. In terms of the 
local coordinates $u^\pm$, we have
\be
{\cal K}_n = {\cal K}_n (u_\pm) \frac{\partial}{\partial u_\pm}
\;,
\ee
with ${\cal K}_n(u_\pm) = - {\cal K}_n(-u_\pm)$; therefore these vector 
fields generate `parity preserving' holomorphic reparametrizations 
in a neighborhood of $u_\pm =0$. In particular, we have
\be
{\cal K}_1 = \pm 2u_\pm \frac{\partial}{\partial u_\pm}
\;,
\ee
i.e.~${\cal K}_1$ acts as a dilatation operator. In terms of local 
coordinates $u_\pm$ the `linear system' is (see \cite{JulNic96} 
for details)
\be
Y^{-1} \partial_\pm Y = \frac12 \rho^{-1}\partial_\pm \rho
\left(u_\pm - \frac1{u_\pm}\right) \frac{\partial}{\partial u_\pm}
\;,
\ee
and takes values in the Lie algebra of vector fields generated by 
$\{ {\cal K}_n \, | \, n\in\N \}$. The ensuing compatiblity condition 
is just the equation of motion for $\rho$.

We now define the groups $\Eneun$ and $\KE$, respectively, in terms of maps 
from $\Sigma$ and $\tilde\Sigma$ into the complexified group ${\rm E}_8(\C)$.
In order to ensure that the solutions of the equations of motion
are real, we have to impose certain reality constraints, which in 
turn requires that $\Sigma$ and $\tilde\Sigma$ admit a generalized 
`complex conjugation', i.e.~a reflection $r$ such that the points of 
$\Sigma$ and $\tilde\Sigma$ invariant under $r$ define `real sections' 
$R(\Sigma)\subset\Sigma$ and $R(\tilde\Sigma)\subset\tilde\Sigma$, 
respectively. We furthermore demand that this reflection commute 
with the involution $\cI$
\be
r \circ \cI = \cI \circ r
\;.
\ee
Of course, not every Riemann surface may admit such a reflection.
Examples of surfaces which do are those surfaces which can be realized
as multisheeted coverings of the complex plane, such that the set
of branch cuts is invariant w.r.t.~reflection on the real axis;
the reflection $r$ is then just the one induced by ordinary
complex conjugation. In particular, if $\Sigma$ is the Riemann 
sphere, we have $R(\Sigma) = \R$. The `moving' branch cut is also
constrained and must be invariant under $r$: for stationary 
axisymmetric solutions, it lies on the imaginary axis and is symmetric
w.r.t.~the real axis, while for Lorentzian solutions, it is a part
of the real axis, or of $R(\Sigma)$ for more general $\Sigma$.

The group $\Eneun$ is defined as 
\be
\Eneun := \Big\{ G\in \CM\, (\Sigma, {\rm E}_8(\C)) \Big|
G(r(w))= \overline{G(w)} \; ; \;
G(w) \in \EE \; \mbox{for} \; r(w)=w  \Big\}
\;,
\ee
where $\CM$ denotes the {\em meromorphic} maps. That is, $\Eneun$ consists 
of meromorphic mappings of $\Sigma$ into the complexified group 
${\rm E}_8(\C)$ with the additional reality constraint that on 
the real section of $\Sigma$ the elements belong to the particular 
real form $\EE$. The above definition generalizes the corresponding 
one of \cite{BreMai87} for stationary axisymmetric solutions 
(where ${\rm G}={\rm SL}(2,\C)$), and can be shown to act transitively 
on the multi-soliton solutions of \cite{BelZak78,BreMai87}.

Similarly, $\KE$ is defined as the group of meromorphic mappings 
\be
\KE \subset \CM \, (\tilde\Sigma, {\rm E}_8(\C))
\;,
\ee
subject to the three requirements
\begin{enumerate}
\item $H(r(\sr)) = \overline{H (\sr)}\;$ for all $\sr\in\tilde\Sigma$.
\item For all $H\in \KE$
\be
H^{-1} = \tau \circ H \circ \cI
\;,
\ee 
where $\tau$ is the symmetric space involution defining the real form 
${\rm E}_{8(8)}$. In particular, we have $H\in {\rm SO}(16)$ 
at the two fixed 
points of $\cI$.
\item All $H\in\KE$ are {\em holomorphic at the branch points}, 
i.e.~the fixed points of $\cI$. 
\end{enumerate}

The first of these conditions is the reality constraint. The second 
is a restatement of the condition \Ref{KE9}; it restricts the affine
Lie algebra to a maximal subalgebra without central extension.
The third (regularity) requirement will be motivated below.

Next we spell out what these conditions imply for the Lie algebra 
$\mathfrak{ke}_{9}$. For an element $h\in\mathfrak{ke}_{9}$, we have
\ba\label{ke9}
h (\sr) &=&  - h^T\left(\frac1{\sr}\right) \quad \Longrightarrow\nn\\ 
h (\sr) &=& \ft12 h_0^{IJ} X^{IJ} +
\sum_{n=1}^\infty \Big[\ft12\left(\gamma^{-n}\!+\!\gamma^n \right) 
h^{IJ}_{n}\, X^{IJ}
+\left(\gamma^{-n}\!-\!\gamma^n \right) h^{A}_{n}\, Y^{A}\Big] 
\;.
\ea
The reality constraint (1) above then implies that the coefficients
$h^{IJ}_n$ and $h^A_n$ are {\em real}. Furthermore, expressing the 
nine Chevalley generators of $\mathfrak{ke}_{9}$ in terms  of the 
current algebra representation above, it is an elementary exercise
(which we leave to the reader) to check that the parametrization~\Ref{ke9} 
is equivalent to the abstract definition~\Ref{Inv} of $\ke$ given in 
the introduction. The sum on the r.h.s. can be 
treated as a formal power series in $\sr$, but it is often more 
convenient to work directly with the condition on the function $h(\sr)$.
The expansion shows that the involutory subalgebra $\ke$ is itself 
{\em not} a Kac Moody algebra; for instance, it does not possess a 
triangular decomposition. It is also easy to see that the central 
term of $\eneun$ does not belong to the subalgebra $\ke$, so the latter
is a maximal `anomaly-free' subalgebra of $\eneun$.

Because $h(\sr)$ is holomorphic at the branch points $\sr = \pm1$
by assumption, it can be expanded into convergent series about the two
branch points in terms of the local coordinates $u_\pm$, viz.
\be
h^\pm (u_\pm) =  \ft12 h^{\pm IJ} (u_\pm) X^{IJ} + h^{\pm A} (u_\pm) Y^A 
\;,
\ee
with 
\be\label{hconstraints}
h^{\pm IJ} (u_\pm) = h^{\pm IJ}(-u_\pm) \;, \quad 
h^{\pm A} (u_\pm) = - h^{\pm A} (-u_\pm)
\;,
\ee
where the superscripts on $h^\pm$ are to indicate that we are
dealing with two expansions of the {\em same} function $h$ about
the two different points. Hence,
\be\label{hpm}
h^\pm (u_\pm) = \ft12 h_0^{\pm IJ} X^{IJ} +
\sum_{n=1}^\infty \Big[\ft12 u_\pm^{2n} h^{\pm IJ}_{2n}\, X^{IJ} + 
u_\pm^{2n-1} h^{\pm A}_{2n-1}\, Y^{A}\Big]
\;.
\ee
The expansions in terms of the coefficients $h_n^\pm$ cannot
terminate at finite order for either $u_+$ or $u_-$, because otherwise 
$h$ would blow up at the opposite branch points contrary to assumption. 
We remark that $\mathfrak{so}(16)\oplus\mathfrak{so}(16)$ is {\em not} 
a subalgebra of $\ke$, despite the occurrence of two expansion coefficients 
$h_0^{\pm IJ}$, since both expansions arise from a single function 
$h(\sr)$. When expanded about the two branch points, the algebra 
$\ke$ looks like `half' of a twisted version of $\eneun$, i.e.~like 
a Borel subalgebra thereof. However, we cannot extend it to a full 
twisted affine Lie algebra because the required holomorphicity 
of $h$ at $u_\pm =0$ eliminates `one half' of the latter.

\mathon
\section{Some remarks on bosonic representations of $\KE$}
\mathoff

It is rather straightforward to construct bosonic (i.e.~single
valued) representations of $\KE$ by `lifting' representations 
of ${\rm E}_{8(8)}$. The more difficult task, however, is to 
come up with {\em irreducible} representations \cite{PreSeg86}. 
The representation which is perhaps most easily understood is the 
{\em adjoint representation}: it is realized in terms of meromorphic 
functions $\{\phi^{IJ}(\sr), \phi^A(\sr)\}$ satisfying the same 
constraints as \Ref{hconstraints} when expanded in local coordinates
about $u_\pm =0$, i.e.
\be\label{phiconstraints}
\phi_\pm^{IJ} (u_\pm) = \phi_\pm^{IJ}(-u_\pm) \;\; , \quad
\phi_\pm^A (u_\pm) = - \phi_\pm^A (-u_\pm)
\;,
\ee
(note that $\phi$ is allowed to have poles of any given order
at $u_\pm =0$). The transformations 
\ba\label{KE9adjoint}
\delta \phi_\pm^{IJ}(u_\pm) &=& 
2h^{\pm K[I}(u_\pm) \phi_\pm^{J]K} (u_\pm)+ 
\ft12 h^{\pm A}(u_\pm) \Gamma^{IJ}_{AB} \phi_\pm^B (u_\pm) \;,
\nn\\
\delta \phi_\pm^A(u_\pm) &=& \ft14 h^{\pm IJ}(u_\pm) 
\Gamma^{IJ}_{AB} \phi_\pm^B (u_\pm) +
\ft14 h^{\pm B}(u_\pm) \Gamma^{IJ}_{BA} \phi_\pm^{IJ} (u_\pm)
\;,
\ea
then preserve \Ref{phiconstraints}. The commutator of two such 
transformations with parameters $h_1$ and $h_2$ is a new 
transformation with parameter
\ba
\tilde{h}^{\pm IJ} (u_\pm)&=& [h_1^\pm (u_\pm), h_2^\pm (u_\pm) ]^{IJ} 
 + \ft12 h_1^{\pm A}(u_\pm) \Gamma^{IJ}_{AB} h_2^{\pm B}(u_\pm) \;,
 \nn\\[.5ex]
\tilde{h}^{\pm A}(u_\pm) &=& \ft12 h_1^{\pm B}(u_\pm) 
h_2^{\pm KL} (u_\pm)\Gamma^{KL}_{BA}
\;.
\ea
Due to the required regularity and the parity constraints 
\Ref{hconstraints}, the second term on the r.h.s. of the first 
equation starts only at ${\cal O}(u_\pm^2)$. The full symmetry 
acting on this, and in fact any other, representation of $\KE$ 
is the semi-direct product of $\KE$ and the restricted 
diffeomorphisms \Ref{Kn}, which is a maximal `anomaly free' 
subalgebra of the semidirect product of the Witt-Virasoro 
(pseudo)group and ${\rm E}_{9(9)}$ preserving the parity 
conditions \Ref{phiconstraints}.

In an analogous fashion, any representation of ${\rm E}_{8(8)}$ can 
be `lifted' to produce a representation of $\KE$ and its semidirect
product with the involutory diffeomorphisms, if suitable `parity 
conditions' analogous to \Ref{phiconstraints} are imposed.
In contradistinction to the Lie algebra $\ke$ itself, its 
representations may have poles of bounded (but arbitrary) 
order at $u_\pm =0$; the holomorphicity requirement ensures that 
the order of the pole is unchanged under the action of $\ke$, 
and also preserved under the diffeomorphisms generated by the vector 
fields ${\cal K}_n$ by virtue of the assumed regularity around 
$u_\pm =0$. The requirement that the representations be complex 
analytic functions is essential here. If we were dealing with arbitrary 
functions instead, these representations would be highly reducible: 
a smaller representation can always be obtained restricting to the 
set of functions which vanish on an arbitrary closed set~\cite{PreSeg86}. 
However, this way of reducing a given representation obviously 
does not work for analytic functions. 

The linear systems \Ref{Lbos} and \Ref{LS1} transform both
as $\ke$ {\em gauge connections}. To see this in more detail, 
observe that the r.h.s. of the linear systems \Ref{Lbos}
and \Ref{LS1} both belong to $\ke$ --- in contrast to the Cartan 
form \Ref{VdV}, which has components in all of the Lie algebra 
$\eee$ (and not just its compact subalgebra $\mathfrak{so}(16)$).
For instance, under \Ref{E9KE9}, the bosonic linear system
\ba\label{LS2}
\hCV^{-1}\partial_\pm \hCV(u_\pm) &=& 
Q_\pm  + u_\pm^{-1} \,P_\pm \;\;\; ;
\ea
is inert under rigid $\Eneun$ (i.e.~does not transform under $G(w)$), and 
transforms with an {\em inhomogeneous term} under $\KE$; the same is
true for the linear system with fermions \Ref{LS1}). However, a 
general $\KE$ gauge transformation will not preserve the particular 
$u_\pm$ dependence on the r.h.s.~of \Ref{LS2}. This is only the case 
for very special $\KE$ transformations with parameters depending on 
the physical fields in a special way --- leading to the non-linear and 
non-local realization of the affine symmetry on the finitely many 
physical fields that is known from the realization of 
the Geroch group in general relativity \cite{DieHoe84,BreMai87}. 
Because they leave the equations of motion form invariant, we will 
refer to these restricted $\KE$ transformations as `on shell'. 

To allow for the most general `off shell' $\KE$ gauge transformation, 
we relax the special $u_\pm$ dependence of \Ref{LS2} by writing
\footnote{We will not always indicate the $x$-dependence in 
the remainder.} 
\ba\label{VdV3}
\hCV^{-1}\partial_\pm \hCV(x;\sr) &=&
\CQ_\pm(x;\sr) + \CP_\pm(x;\sr)~\equiv~ \CJ_\pm(x;\sr)
= - \CJ^T_\pm \left(x; \frac1{\sr}\right) \in\ke \;.
\ea
The `dual potentials', which were previously constrained by 
the special $u_\pm$ dependence on the r.h.s.~of \Ref{Lbos} 
(or \Ref{LS1}) to be non-linear and non-local functions of the 
physical fields now become independent gauge degrees of freedom. 
The currents $\CJ_\pm$ can be expanded about both branch points,
such that
\ba\label{VdV4}
\CJ_\pm \Big|_{\sr \,\sim \, \mp 1} &=& \CQ_\pm (u_\pm) + \CP_\pm( u_\pm)
\;,
\ea
with
\ba
\CQ_\pm(u_\pm) &=& \CQ_\pm(- u_\pm) = Q_\pm + \CO(u_\pm^2) \;,\non
\CP_\pm(u_\pm) &=& - \CP_\pm(-u_\pm) =
u_\pm^{-1} \,P_\pm + \CO(u_\pm) \;,
\label{QP0}
\ea
where we only require that $\CP_\pm$ has at most a first order 
pole, and regular $Q_\pm$. At the opposite branch point we demand
\ba\label{VdV5}
\CJ_\pm \Big|_{\sr \, \sim \, \pm 1} &=& \tilde\CQ_\pm (u_\mp) + 
\tilde\CP_\pm( u_\mp)
\;,
\ea
with
\ba
\tilde\CQ_\pm(u_\mp) &=& \tilde\CQ_\pm(- u_\mp) = 
\tilde Q_\pm + \CO(u_\mp^2) \;,\non
\tilde\CP_\pm(u_\mp) &=& - \tilde\CP_\pm(-u_\mp) =
u_\mp \,\tilde P_\pm + \CO(u_\mp^3) \;.
\label{QPt0}
\ea
While $\CP_\pm$ thus has a first order pole at $u_\pm =0$, it has a 
zero at the opposite branch point $u_\pm = \infty$, and this property 
is preserved by the transformations \Ref{dJ}. Analogous comments apply 
to the linear system with fermions, except that now $\CP_\pm (u_\pm) 
\sim u_\pm^{-3}$ and $\CQ_\pm (u_\pm) \sim u_\pm^{-4}$ for $u_\pm \sim 0$.
It is only for the gauge-fixed form of the linear system \Ref{LS2}
that $Q_\pm=\tilde Q_\pm$ and $P_\pm=\tilde P_\pm$.

Under $\delta\hCV=\hCV\,h$, the connection $\CJ_\pm (\sr)$ 
transforms according to
\ba\label{dJ}
\delta \CJ_\pm(\sr) &=& \partial_\pm h (\sr) 
+ \left[\CJ_\pm(\sr) ,h (\sr)\right] \;.
\ea
In terms of the coordinates $u_\pm$ and making use of \Ref{spdiff1} 
we have
\ba\label{dJ1}
\delta \CJ_\pm(u_\pm) &=& \partial_\pm h^\pm (u_\pm) +
\ft12(u_\pm-u_\pm^{-1})\,\rho^{-1}\partial_\pm\rho \, \partial h^\pm (u_\pm) 
+ \left[\CJ_\pm(u_\pm) ,h^\pm (u_\pm)\right]
\;,
\ea
where the derivative in the first term does not act on $u_\pm$.
\Ref{dJ} implies
\ba\label{dPdQ}
\delta \CQ_\pm^{IJ} (u_\pm) &=& \partial_\pm h^{\pm IJ} (u_\pm) +
\ft12(u_\pm-u_\pm^{-1})\,\rho^{-1}\partial_\pm\rho 
\, \partial h^{\pm IJ} (u_\pm) 
\non
&&{}+
   2 \CQ_\pm^{K[I} (u_\pm) h^{\pm J]K} (u_\pm) + 
   \ft12 \CP_\pm^A(u_\pm) \Gamma^{IJ}_{AB} h^{\pm B}(u_\pm)\;, 
\non[1ex]
\delta \CP_\pm^{A} (u_\pm) &=& \partial_\pm h^{\pm A} (u_\pm) +
\ft12(u_\pm-u_\pm^{-1})\,\rho^{-1}\partial_\pm\rho 
\, \partial h^{\pm A} (u_\pm) 
\non
&&{}+
   \ft14 \CQ_\pm^{IJ} (u_\pm) \Gamma^{IJ}_{AB} h^{\pm B} (u_\pm) + 
   \ft14 \CP_\pm^B(u_\pm) \Gamma^{IJ}_{BA} h^{\pm IJ}(u_\pm) \;.
\ea
Substituting \Ref{hpm}, we obtain for instance
\be\label{dPA}
\delta P^A_\pm = -\ft12 \rho^{-1} \partial_\pm \rho \, h_1^{\pm A}
             - \ft14 P_\pm^B \Gamma^{IJ}_{AB} h_0^{\pm IJ} 
\;.
\ee

The key point here is that the information about the physical fields
(the currents $P_\pm$ and the fermionic bilinears) {\em is encoded in
the poles of the} $\KE$ {\em connection}, whereas the higher order 
regular terms in $u_\pm$ in the expansion of $\CJ_\pm (u_\pm)$ are to be 
interpreted as $\KE$ gauge degrees of freedom, which can be removed 
by the adjoint action of $\KE$. The residue is only affected by the 
inhomogeneous term in \Ref{dJ}, and it is precisely this term which 
allows to generate non-trivial solutions from the vacuum by the nonlinear 
and non-local action of $\Eneun$ via the induced action of $\KE$.

\mathon
\section{Induced action of $\KE$ on the supergravity multiplet}
\mathoff

In our previous work \cite{NicSam98}, we have studied the action of 
the infinite dimensional global $\Eneun$ symmetry on all fields
of $N=16$ supergravity. On the chiral bosonic `currents' and on
the chiral components of the fermionic fields, the global $\Eneun$
was shown to act via an induced $\KE$ transformation, which can be
canonically generated via a Lie-Poisson action. We refer readers 
to \cite{NicSam98} for the detailed derivations and explicit expressions, 
and here only summarize the pertinent formulas for the variations of 
the various $N=16$ supergravity fields. If the action of the global
$\Eneun$ is parametrized by a Lie-algebra-valued function 
$\Lambda(w)\!\in\!\eee$, the solution $\hCV$ of the linear system 
transforms as
\be\label{actionVh}
\delta_\Lambda\,\hCV(x,\sr(w)) ~=~
\oint_{\CC}\,\frac{dv}{2\pi\I\,(v\!-\!w)}\,
\Lambda(v) \hCV(x,\sr(w)) -
\hCV(x,\sr(w))\,\Upsilon_{\!\Lambda}(x,\sr(w))\;, 
\ee
with the matrix
\ba
\Upsilon_{\!\Lambda}(x,\sr(w))&\equiv&
\oint_{\CC}\:
\frac{d\sr'}{2\pi\I\,\sr'}\,
\frac{\sr(w)(1-\sr'{}^2)}{(\sr'\!-\!\sr(w))\,(1\!-\!\sr(w)\sr')}
\;
\ft12\hL^{IJ}(\sr') \,X^{IJ} \non[1ex]
&&{}+
\oint_{\CC}\:
\frac{d\sr'}{2\pi\I\,\sr'}\,
\frac{\sr'(1\!-\!\sr(w)^2)}{(\sr'\!-\!\sr(w))\,(1\!-\!\sr(w)\sr')}
\;
\hL^A(\sr')\,Y^A \;,\label{Y}
\ea
and the dressed parameters
\ba\label{dressed}
\hL(\sr') ~\equiv~ 
\ft12 \hL^{IJ}(\sr')\,  X^{IJ} + \hL^A(\sr')\, Y^A
&\equiv&
\hCV^{-1}(\sr')\Lambda(w(\sr')) \hCV(\sr')
\;.
\ea
Replacing $\sr \ra 1/\sr$ in \Ref{Y} one directly verifies that 
$\Upsilon_{\!\Lambda}(x,\sr(w))\in\mathfrak{ke}_{9}$ according to 
\Ref{ke9}, independently of the choice of integration contout $\CC$.
In order to recover the affine (loop) algebra, choose 
$\Lambda(w)=\Lambda_n w^n$ and a path $\CC$ encircling the point 
$\gamma=0$ in the complex $\gamma$-plane. Expanding $\hL$ around 
$\gamma=0$ into its singular and regular part, we get
\ba
\hCV^{-1} \Lambda \hCV ~\equiv~ 
\hL_{\rm sing}  + \hL_{\rm reg}\;,\qquad
\hL_{\rm sing} &\equiv& \sum_{k=0}^n \gamma^{-k}\hL_{-k}   \;.
\ea
Formula \Ref{Y} then yields
\ba
\Upsilon_{\! \Lambda}(x,\sr)&=& 
\sum_{k=0}^n
\ft12\left(\gamma^{-k}\!+\!\gamma^k \right) \hL^{IJ}_{-k}\, X^{IJ}
+\left(\gamma^{-k}\!-\!\gamma^k \right) \hL^{A}_{-k}\, Y^{A}
\non[1ex]
&=& \hL_{\rm sing}(\gamma) -\hL_{\rm sing}^T(1/\gamma) 
\;,
\nn
\ea
and thus
\ba
\delta_\Lambda \,\hCV(\gamma) &=&
\hCV(\gamma)\,\Big( \hL_{\rm reg}(\gamma)+\hL_{\rm sing}^T(1/\gamma) \Big)
\;,
\ea
(note that $\hL_{\rm sing}^T(1/\gamma)$ is regular near $\sr =0$).

These considerations show that our definition \Ref{Y} selects 
$\Upsilon_{\Lambda}(\sr)$ precisely such that $\hCV$ remains 
holomorphic around $\gamma=0$, in accordance with the prescription 
of \cite{BreMai87}. Below we will spell out in more detail the conditions 
that the expansion coefficients of these `on shell $\KE$ transformations'
have to obey, but let us note already here that \Ref{dressed} and the 
linear system together imply the differential identities
\be\label{dressed1}
D_\pm \hL^{IJ}=\frac1{2 u_\pm} \Gamma^{IJ}_{AB}\hL^AP_\pm^B \;, \qquad
D_\pm \hL^{A}=\frac1{4u_\pm} \Gamma^{IJ}_{AB}\hL^{IJ} P_\pm^B \;,
\ee
which express the dependence of the dressed parameters on the
physical fields.

Working out the action of $\delta_\Lambda$ on the various fields,
we obtain the explicit formulas for the bosonic fields~\cite{NicSam98}
\ba\label{E9bosons0}
\delta_\Lambda\, \CV(x) &=& \int_\CC \frac{dv}{2\pi\I}\; \left(
\frac{2\sr}{\rho(1-\sr^2)}\,\CV(x)
\hL^A Y^A \right)
\;,\non[1ex]
\delta_\Lambda\, P_\pm(x) &=& \int_\CC\, \frac{dv}{2\pi\I}\;
\left[\frac{\sr}{\rho(1\pm\sr)^2}\,
\hL^{IJ} X^{IJ}
 ,\, P_\pm(x)\right] \non
&&{}\hspace*{11em}\mp\int_\CC\, \frac{dv}{2\pi\I}\;
\frac{4\sr^2\partial_\pm\rho} {\rho^2(1\pm\sr)^2(1-\sr^2)}\; 
\hL^A Y^A\;,
\ea
and the fermionic fields\footnote{ Note the change w.r.t.~the formulas 
in \cite{NicSam98} by a relative factor of $-1/2$.}
\ba\label{E9fermions0}
\delta_\Lambda\,  \psi_{2\,\pm}^{I} &=& 
 \oint_\CC\frac{dv}{2\pi \I }\,\left(\frac{2\sr}{\rho(1\!\pm\!\sr)^2}\,
\hL^{IJ}(\sr)\psi_{2\,\pm}^J \right)\;,\non[1ex]
\delta_\Lambda\,  \chi_\pm^{\dot{A}} &=& 
\oint_\CC\frac{dv}{2\pi \I }\,\left(\frac{\sr}{2\rho(1\!\pm\!\sr)^2}\,
\Gamma^{IJ}_{\dot{A}\dot{B}}\,\hL^{IJ}(\sr)\chi^{\dot{B}}_\pm +
\frac{4\sr^2}{\rho(1\!\pm\!\sr)^2(1\!-\!\sr^2)}\,
\Gamma^I_{A\dot{A}}\,\hL^{A}(\sr)\psi_{2\,\pm}^I
\right),\non[1ex]
\delta_\Lambda\,  \psi_{\pm}^{I} &=&
\oint_\CC\frac{dv}{2\pi \I }\,\left(\frac{2\sr}{\rho(1\!\pm\!\sr)^2}\,
\hL^{IJ}(\sr)\psi_{\pm}^J 
+\frac{8\sr^2}{\rho(1\!\pm\!\sr)^4}\,
\hL^{IJ}(\sr)\psi_{2\,\pm}^J \right)\non
&& \mp  \oint_\CC\frac{dv}{2\pi i }\,\left(
\frac{4\sr^2}{\rho(1\!\pm\!\sr)^2(1\!-\!\sr^2)}\,
\Gamma^I_{A\dot{B}}\,\hL^{A}(\sr) \chi^{\dot{B}}_\pm \right)\;.
\ea
Writing 
\ba
\Upsilon_{\!\Lambda}(x,\sr(w)) &\equiv &
\ft12\,\Upsilon^{IJ}\,X^{IJ} + \Upsilon^{A}\,Y^{A}
\;,
\ea
making use of the formula (cf. \Ref{gdg})
\be
\frac{d\sr}{\sr} = - \frac{2\sr}{\rho (1-\sr^2)} dw
\;,
\ee
and recalling the definition \Ref{Y}, we see that the symmetry action 
of $\KE$ on the bosonic physical fields may be written in a 
much simpler form as
\ba\label{E9bosons}
\delta_\Lambda\, \CV &=& -\CV \,\Upsilon_{\!\Lambda}|_{\sr=0} 
\;,\non[1ex]
\delta_\Lambda\, P^A_\pm&=&
\ft14\,\Gamma^{IJ}_{AB}\,P^B_\pm\,
\Upsilon^{IJ}{}|_{\sr=\mp1}\,
~\pm~
\rho^{-1}\partial_\pm\rho\, \partial_\sr \Upsilon^{A}{}\big|_{\sr=\mp1} 
\;,
\non[1ex]
\delta_\Lambda\,\sigma &=& \oint_\CC\, \frac{d\sr}{2\pi\I}\;
\tr\Big[ \Lambda\partial_\sr\hCV\hCV^{-1}\Big]\;.
\ea
The action on the fermionic fields of the model takes the form
\ba \label{E9fermions}
\delta_\Lambda\, \psi_{2\,\pm}^{I} &=& 
\psi_{2\,\pm}^J \,
\Upsilon^{IJ}{}|_{\sr=\mp1}\;,\non[1ex]
\delta_\Lambda\, \chi_\pm^{\dot{A}} &=& 
\ft14\Gamma^{IJ}_{\dot{A}\dot{B}}\,\chi^{\dot{B}}_\pm\,
\Upsilon^{IJ}{}|_{\sr=\mp1}
- \Gamma^I_{A\dot{A}}\,\psi_{2\,\pm}^I \,
\partial_\sr\Upsilon^{A}\big|_{\sr=\mp1} \;, \non[1ex]
\delta_\Lambda\, \psi_{\pm}^{I} &=&
 \psi_{\pm}^J \,
\Upsilon^{IJ}{}|_{\sr=\mp1}
\pm \Gamma^I_{A\dot{B}}\,\chi^{\dot{B}}_\pm\,
\partial_\sr\Upsilon^{A}\big|_{\sr=\mp1}
\mp2\psi_{2\,\pm}^J \,
\big(\partial_\sr^2\Upsilon^{IJ}\mp\partial_\sr\Upsilon^{IJ}\big)_{\sr=\mp1}  \;.
\ea
Switching to local coordinates $u_\pm$, we have the expansion coefficients
$\{ \Upsilon^{\pm IJ}, \Upsilon^{\pm A} \}$ about the branch points
\ba\label{Ups}
\Upsilon^\pm_{\!\Lambda}(x,u_\pm) &\equiv &
\sum_{n=0}^\infty \left(
\ft12\,u_\pm^{2n}\, \Upsilon_{2n}^{\pm IJ}\,X^{IJ} +
 u_\pm^{2n+1}\, \Upsilon_{2n+1}^{\pm A}\,Y^{A}
\right)
\;,\qquad \Xi^A_0 ~\equiv~ \Upsilon^A\big|_{\gamma=0}\;.
\ea
They satisfy differential equations which follow from \Ref{Y} 
and \Ref{dressed1}, namely 
\ba
D_\pm \Upsilon^{\mp IJ}_{2n} &=&
\ft12\,\Gamma^{IJ}_{AB} P_\pm^B\,\Upsilon^{\mp A}_{2n-1}
+ \left( (1-n)\Upsilon^{\mp IJ}_{2n-2}+ n \Upsilon^{\mp IJ}_{2n} 
\right)\rho^{-1}\partial_\pm\rho
\;,\qquad (n\ge1)
\non[1ex]
D_\pm \Upsilon^{\mp IJ}_{0} &=&
-\ft12\,\Gamma^{IJ}_{AB} P_\pm^B\,\Xi^{A}_{0}
\;,\non[1ex]
D_\pm \Upsilon^{\mp A}_{2n+1} &=&
\ft14\,\Gamma^{IJ}_{AB} P_\pm^B\,\Upsilon^{\mp IJ}_{2n}
+ \left((n\!+\!\ft12) \Upsilon^{\mp A}_{2n+1} + 
(\ft12\!-\!n)\Upsilon^{\mp A}_{2n-1} \right)\rho^{-1}\partial_\pm\rho
\;,\qquad (n\ge1)
\non[1ex]
D_\pm \Upsilon^{\mp A}_{1} &=&
\ft14\,\Gamma^{IJ}_{AB} P_\pm^B
\left(\Upsilon^{\mp IJ}_{0}-\Upsilon^{\pm IJ}_{0} \right)
+ \ft12\left(\Upsilon^{\mp A}_{1} - \Upsilon^{\pm A}_{1}
\right)\rho^{-1}\partial_\pm\rho
\;,
\label{diffY}
\ea
and similarly
\ba
D_\pm \Upsilon^{\pm IJ}_{2n} &=&
\ft12\,\Gamma^{IJ}_{AB} P_\pm^B\,\Upsilon^{\pm A}_{2n+1}
+ \left((1+n)\Upsilon^{\pm IJ}_{2n+2} - n \Upsilon^{\pm IJ}_{2n} 
\right)\rho^{-1}\partial_\pm\rho
\;,\qquad (n\ge1)
\non[1ex]
D_\pm \Upsilon^{\pm IJ}_{0} &=&
\ft12\,\Gamma^{IJ}_{AB} P_\pm^B\,\left(\Upsilon^{\pm A}_1-\Xi^{A}_{0}\right)
+ \Upsilon^{\pm IJ}_{2}\,\rho^{-1}\partial_\pm\rho
\;,\non[1ex]
D_\pm \Upsilon^{\pm A}_{2n+1} &=&
\ft14\,\Gamma^{IJ}_{AB} P_\pm^B\,\Upsilon^{\pm IJ}_{2n}
+ \left((n+\ft32) \Upsilon^{\pm A}_{2n+3} - (n+\ft12)\Upsilon^{\pm A}_{2n+1}
\right)\rho^{-1}\partial_\pm\rho
\;.
\label{diffY2}
\ea
Note that only the relations for $\Upsilon^{\pm IJ}_{0}$, 
$\Upsilon^{\pm A}_{1}$ are not chiral. 

It is then convenient to express the transformation of the
fields~\Ref{E9bosons}, \Ref{E9fermions} in terms of these expansion
coefficients. For the bosons we obtain
\ba\label{E9ff}
\delta_\Lambda\,\CV &=& -\CV\,\Xi_0^A \,Y^A \;,\qquad
 \delta_\Lambda\,Q^{IJ}_\pm ~=~ \ft12\,\Gamma^{IJ}_{AB}\,
 \Xi^A_0\,P_\pm^B \;,
\non[1ex]
\delta_\Lambda\, P^A_\pm&=&
\ft14\,\Gamma^{IJ}_{AB}\,P^B_\pm\,
\Upsilon_0^{\pm IJ}{}\,
~+~
\ft12\,\rho^{-1}\partial_\pm\rho\, \Upsilon^{\pm A}_1
\;,
\non[1ex]
\delta_\Lambda\, (\partial_\pm\sigma)  &=& 
\ft12\,P_\pm^A\,\Upsilon^{\pm A}_1  
\;.
\ea
The second of these formulas tells us that $P^A_\pm$ indeed
transforms as a component of a $\KE$ connection, cf. \Ref{dPA};
the same can be verified for $Q_\pm^{IJ}$ by use of the second 
relation in \Ref{diffY2}. The transformation for the derivative 
of the conformal  factor is in accord with the conformal 
constraints~\Ref{TTT}. For the fermions, we get
\ba\label{E9ff1}
\delta_\Lambda\, \psi_{2\,\pm}^{I} &=& 
\psi_{2\,\pm}^J \,
\Upsilon_0^{\pm IJ}{}\;,\non[1ex]
\delta_\Lambda\, \chi_\pm^{\dot{A}} &=& 
\ft14\Gamma^{IJ}_{\dot{A}\dot{B}}\,\chi^{\dot{B}}_\pm\,
\Upsilon_0^{\pm IJ}
\mp\ft12\,\Gamma^I_{A\dot{A}}\,\psi_{2\,\pm}^I \,
\Upsilon^{\pm A}_1\;, \non[1ex]
\delta_\Lambda\, \psi_{\pm}^{I} &=&
 \psi_{\pm}^J \,
\Upsilon_0^{\pm IJ}
+\ft12\, \Gamma^I_{A\dot{B}}\,\chi^{\dot{B}}_\pm\,
\Upsilon_1^{\pm A}
\mp\psi_{2\,\pm}^J \,
\Upsilon^{\pm IJ}_2\;.
\ea
The differential relations \Ref{diffY}, \Ref{diffY2} are crucial to
check the invariance of the equations of motion under these
transformations.

We have thus realized $\KE$ as a group of symmetry transformations 
on finitely many fields in such a way that the variations of the 
chiral components of the bosons and fermions are expressed in
terms of the lowest coefficients of the transformation parameter 
$h(\sr)$ about the two branch points $\sr = \pm 1$. This representation
of $\KE$, however, is {\em not} faithful, because the variations are only
sensitive to the terms up to second order. This is even more evident 
for the redefined fermion fields
\ba
\widetilde{\psi}_{2\pm}^I &\equiv& \psi_{2\pm}^I 
\;,
\non[1ex]
\widetilde{\chi}_\pm^\dA &\equiv& 
\chi_\pm^\dA \mp \frac1{\rho^{-1}\partial_\pm\rho}\,
\Gamma^I_{A\dA} P_\pm^A \widetilde{\psi}_{2\pm}^I 
\;,
\non[1ex]
\widetilde{\psi}_\pm^I &\equiv& \psi_\pm^I 
\pm \frac1{\rho^{-1}\partial_\pm\rho}\,
\left(
\rho^{-1}\,D_\pm(\rho\widetilde{\psi}_{2\pm}^I) + 
\partial_\pm\sigma\widetilde{\psi}_{2\pm}^I 
\mp \Gamma^I_{A\dA}P_\pm^A\widetilde{\chi}^\dA_\pm
\right)
\;.
\ea
It is straightforward to check that these new fermion fields 
transform separately under $\KE$ and `see' only the zero modes:
\ba
\delta_\Lambda\, \widetilde{\psi}_{2\,\pm}^{I} = 
\Upsilon_0^{\pm IJ}\,\widetilde{\psi}_{2\,\pm}^J \,
{}\;,\qquad
\delta_\Lambda\, \widetilde{\chi}_\pm^{\dot{A}} =
\ft14\Gamma^{IJ}_{\dot{A}\dot{B}}\,\Upsilon_0^{\pm IJ}\,
\widetilde{\chi}^{\dot{B}}_\pm
\;, \qquad
\delta_\Lambda\, \widetilde{\psi}_{\pm}^{I} =
 \Upsilon_0^{\pm IJ}\,\widetilde{\psi}_{\pm}^J 
\;.
\label{cf}
\ea
Because $\widetilde{\psi}_\pm^I$ is essentially the supersymmetry
constraint~\Ref{TS} (the change of sign in the $\partial_\pm\sigma$
term is due to the change from $\chi$ to $\widetilde{\chi}$), the last
of these formulas implies a similar formula for the supersymmetry
constraints, viz.
\be\label{dSI}
\delta_\Lambda\, {\cal S}_{\pm}^{I} =
 \Upsilon_0^{\pm IJ}\, {\cal S}_{\pm}^J
 \;,
\ee
in agreement with the results of section 5.3 of \cite{NicSam98}.
This looks like the action of a chiral ${\rm Spin}(16)_+ \times 
{\rm Spin}(16)_-$ symmetry \cite{Nico91}, but, as we have already 
emphasized, the latter is not a subgroup of $\KE$, and hence not 
a symmetry of the reduced theory.

\mathon
\section{A spinor representation of $\KE$ at low orders}
\mathoff

As we have just shown, the contour integral expressions derived 
in \cite{NicSam98} and describing the induced on shell action 
of $\KE$ on all fields, can be brought to the rather more simple 
form given in~\Ref{E9ff} and \Ref{E9ff1}. The crucial point about these 
formulas is that they express the variations in terms of the coefficient 
functions $\Upsilon_n^\pm$ obtained by expanding about the branch points 
$\sr = \pm 1$ as in \Ref{Ups}, and that these expansions are in 
accord with the chiral decomposition of the worldsheet bosons 
and fermions. In this way, the positive and  negative chiralities 
become `attached' to the fixed points of the involution $\cI$.
We now propose to combine all the fermionic fields 
into a single object, an `on shell' $\KE$ spinor made up of the
fields of $N=16$ supergravity. This spinor is expected to be part
of an as yet unknown `off shell' spinorial representation of $\KE$ 
with {\em infinitely many} components. We will argue that the local 
supersymmetry parameters should similarly be enlarged to a spinor 
representation of $\KE$ in an off shell version of the theory.

To this aim, we return to the expansion \Ref{Ups} in terms of local 
coordinates $u_\pm$. In order to combine the corresponding transformations
on the fermions into a single formula, we introduce $u_\pm$ (and, 
of course, $x$-) dependent spinors $\Psi_\pm^I(x,u_\pm)$ and 
$\CX^\dA_\pm (x,u_\pm)$ in the local patches coordinatized by 
$u_\pm$, subject to the parity constraints
\be
\Psi^I_\pm (u_\pm) = \Psi^I_\pm (-u_\pm)  \; , \qquad
\CX^\dA_\pm (u_\pm) = - \CX^\dA_\pm (-u_\pm)
\;,
\ee
and in such a way that the $N=16$ supergravity fields appear as the 
lowest components, viz.
\ba\label{fermions}
\Psi_\pm^I (u_\pm) &=& (u_\pm^{-2}-1) \,\psi_{2\pm}^I 
  ~ \mp~ \psi^I_\pm ~+~ \cO (u_\pm^2)\;, \nn\\[1ex]
\CX^\dA_\pm (u) &=& \mp u_\pm^{-1} \,\chi^\dA ~+~ \cO(u_\pm)\;.
\ea
It is important here that, as for the linear system, but unlike for 
the $\KE$ transformations, we do admit singular terms, and that in 
this way the fermions of $N=16$ supergravity become associated to 
the poles in this expansion, in analogy with the first order 
poles at $u_\pm =0$ multiplying the scalar currents $P_\pm^A$ in 
the bosonic linear system. It is then straightforward to check that 
the transformations \Ref{E9ff1} are recovered from the $\cO(u_\pm^{n})$ 
terms for $n=-2,-1,0\;$ by expanding
\ba\label{SUSYtrafo2}
\delta\Psi_\pm^I (u_\pm) &=& \Upsilon^{IJ}_\pm (u_\pm) \Psi^J_\pm (u_\pm)
   + \ft12\,\Gamma^I_{A\dA} \Upsilon_\pm^A (u_\pm) \CX^\dA_\pm (u_\pm) \;,
   \nn\\[1.5ex]
\delta\CX^\dA_\pm (u_\pm) &=& 
\ft14 \Gamma^{IJ}_{\dA\dB}  \Upsilon^{IJ}_\pm (u_\pm) \CX^\dB_\pm (u_\pm) +
  \ft12\,\Gamma^I_{A\dA} \Upsilon_\pm^A (u_\pm) \Psi_\pm^I (u_\pm)
\;,
\ea
in $u_\pm$, making use of \Ref{Ups} and \Ref{fermions}. Because
the action of $\KE$ mixes gravitinos and matter fermions, one must
consider $\{ \Psi, \CX \}$ as a {\em single object}. 
Furthermore, this action is double-valued because of the 
double-valuedness of the $Spin(16)$ representations $\bf{16}$ and 
$\bf{128}_c$ w.r.t.~to the $Spin(16)$ subgroup of $\KE$ \cite{Keur03}.
For this reason, we are indeed dealing with the beginnings of a 
{\em spinorial representation} of $\KE$.
 
However, \Ref{SUSYtrafo2} cannot be the complete answer, and thus
the $\KE$ spinor must presumably contain further components beyond
those indicated in \Ref{fermions}. The obvious reason for this is the 
failure of the transformations \Ref{SUSYtrafo2} to close into the 
$\ke$ algebra on the components of $\CX^\dA_\pm (u_\pm)$ other than 
the lowest one \footnote{Technically speaking, the main obstacle 
here is the fact that $\Gamma^I \Gamma^{(6)} \Gamma^I = 4 \Gamma^{(6)} 
\neq 0$, whereas $\Gamma^{IJ} \Gamma^{(6)} \Gamma^{IJ}= 0$ implying 
the closure of the transformations \Ref{KE9adjoint} (after an ${\rm SO}(16)$ 
Fierz rearrangement). The corresponding term in the expansion of 
$[\delta_1 , \delta_2] \CX^\dA_\pm (u_\pm)$ is not yet visible at 
$\CO(u_\pm^{-1})$.}. However, they do close properly on all components 
of $\Psi^I_\pm (u_\pm)$, and in particular on all the components 
written out in \Ref{fermions}, i.e.~precisely on the fermions of 
$N=16$ supergravity --- as required by the consistency of the induced 
$\KE$ transformations on the $N=16$ supergravity multiplet. Extra 
fermionic fields will only appear at higher orders in $u_\pm$, but not 
introduce further singular terms (which we would have to associate 
with new fermionic physical degrees of freedom). 

At low orders in $u_\pm$ the ansatz \Ref{fermions} can be corroborated 
in several different ways. First of all, we can show that to the order 
indicated above, the fermionic equations of motion (Eq.~(2.12) 
of~\cite{NicSam98}) can be combined into a single $\KE$ covariant 
`Dirac equation', with the linear system as the $\ke$ gauge connection. 
Namely, to linear order in the fermions, all fermionic equations of motion 
are contained in
\ba\label{Dirac1}
(\CD_\mp \Psi_\pm)^I (u_\pm) &\equiv& \tilde\CD_\mp \Psi^I_\pm (u_\pm) +  
\ft12\Gamma^I_{A\dA} \tilde\CP_\mp^A(u_\pm) \CX^\dA_\pm (u_\pm) ~=~ 0\;, 
\nn\\[1ex]
(\CD_\mp \CX_\pm)^\dA (u_\pm) &\equiv&  \tilde\CD_\mp \CX^\dA_\pm (u_\pm) +  
\ft12\Gamma^I_{A\dA} \tilde\CP_\mp^A(u_\pm) \Psi^I_\pm (u_\pm)~=~ 0\;,
\ea
expanded to lowest orders in $u_\pm$,
where the connection components $\tilde\CQ_\pm^{IJ}$ and $\tilde\CP_\pm^A$
are to be taken from \Ref{QPt0}.
In particular, the terms containing $\rho^{-1}\partial_\pm \rho$
are reproduced correctly from the derivatives acting on $u_\pm$
via the formula \Ref{udu}. The equations of motion tie together the 
expansions in $u_+$ and $u_-$; for instance, the terms proportional
to $P_\mp\chi_\pm$ and $P_\mp \psi_{2\pm}$ have the correct pole order
by virtue of the relation $u_+ u_- =1$. In accordance with~\Ref{SUSYtrafo2} 
and the transformation properties of the linear system, this equation 
is indeed $\KE$ covariant to lowest orders in $u_\pm$.

Evidently the local supersymmetry transformation parameters should 
similarly belong to some spinor representation of $\KE$, such that the 
local supersymmetry parameter of $N=16$ supergravity is but the 
lowest component of an infinite tower of local supersymmetries. 
We denote this representation by $\CE$, and proceed from the hypothesis 
that it is the same as the one for the fermions, except that 
we require $\cal E$ to be regular at $u_\pm =0$. The reason
for this assumption is that this appears to be the only possibility
that will eventually allow us to gauge away all components of 
$\{\Psi^I_\pm, \CX^\dA_\pm \}$ other than the singular ones, which we 
wish to associate with the fermions of $N=16$ supergravity. We are 
thus led to introduce new $x$ and $u_\pm$-dependent spinor parameters
\be
\CE^I_\pm (x,u_\pm) = \CE^I_\pm (x,- u_\pm) \quad , \qquad 
\CE^\dA_\pm (x,u_\pm) = - \CE^\dA_\pm (x,- u_\pm) 
\;,
\ee
which are holomorphic in $u_\pm$ around the branch points
\be
\CE^I_\pm (x,u_\pm) ~=~ \e^I_{0\pm}(x) ~+~ \cO(u_\pm^2) \;, \quad 
\CE^\dA_\pm (x,u_\pm) ~=~ u_\pm \e^\dA_{1\pm}(x) ~+~ \cO(u_\pm^3)
\;.
\label{EEan}
\ee
We postulate that the $N=16$ local supersymmetry parameter $\e^I_\pm$ 
should appear at lowest order in the expansion of $\CE^I_\pm (u_\pm)$, 
such that all other components correspond to new local supersymmetries.
The superconformal constraints on the former \cite{NicSam98} lead us 
to impose the constraint
\ba\label{DE}
(\CD_\mp \CE_\pm)^I (u_\pm) &\equiv& \tilde\CD_\mp \CE^I_\pm (u_\pm) +  
\ft12\Gamma^I_{A\dA} \tilde\CP_\mp^A(u_\pm) \CE^\dA_\pm (u_\pm) ~=~ 0\;, 
\nn\\[1ex]
(\CD_\mp \CE_\pm)^\dA (u_\pm) &\equiv&  \tilde\CD_\mp \CE^\dA_\pm (u_\pm) +  
\ft12\Gamma^I_{A\dA} \tilde\CP_\mp^A(u_\pm) \CE^I_\pm (u_\pm)~=~ 0\;,
\ea
on the generalized supersymmetry parameters. Again, these equations
are $\KE$ covariant in linear order in $u_\pm$. Up to this order, they
are solved by
\ba
\e^I_{0\pm} &\equiv& \e^I_{\pm}\quad \mbox{with $D_\mp\e^I_\pm=0$} \;,
\qquad
\e^\dA_{1\pm}\,\rho^{-1}\partial_\pm \rho ~\equiv~ \Gamma^I_{A\dA}\,
 P^A_{\pm} \,\e^I_{\pm}
\;.
\label{ansatzE}
\ea
The first of these is the expected superconformal constraint
on the $N=16$ supersymmetry parameter. To verify the second
relation for the `new' supersymmetry parameter $\e_\pm^\dA$
requires use of the integrability constraint $D_- P_+^A = D_+ P_-^A$ 
and the equation of motion $D_-(\rho P_+^A) + D_+ (\rho P_-^A) = 0$  
(neglecting cubic fermionic terms). The formula for $\e_\pm^\dA$
can be viewed as resulting from a consistent truncation of an infinite 
number of local supersymmetries to a finite number; the fact that it 
can be expressed in terms of known quantities of $N=16$ supergravity 
again reflects the fact that we are dealing with an `on shell' 
realization of $\KE$ only.

Finally, also the generalized supersymmetry variations can be cast into
the form
\ba
(\delta\Psi)_\pm^I (u_\pm) &=& 
(\CD_\pm \CE_\pm)^I (u_\pm)~\equiv~
   u_\pm^2 \CD_\pm \big(u_\pm^{-2} \CE^I_\pm (u_\pm)\big)
  + \ft12\,\Gamma^I_{A\dA} \CP_\pm^A(u_\pm) \CE^\dA_\pm (u_\pm)\;, \non[1ex]
(\delta \CX)^\dA_\pm (u_\pm) &=& 
(\CD_\pm \CE_\pm)^\dA (u_\pm)~\equiv~
   u_\pm^2 \CD_\pm \big(u_\pm^{-2} \CE^\dA_\pm (u_\pm)\big)
  + \ft12\Gamma^I_{A\dA} \CP_\pm^A(u_\pm) \CE^I_\pm (u_\pm)\;,
  \label{susy33}
\ea
again $\KE$ covariant in lowest orders of $u_\pm$, now with $\CQ_\pm$, $\CP_\pm$ from \Ref{QP0}.
With a little algebra and using formulas \Ref{udu} one shows 
that these formulas yield the correct supersymmetry variations of the 
fermionic fields of $N=16$ supergravity (cf.~Eq.~(2.18) of~\cite{NicSam98}), 
when expanded in powers of $u_\pm$. For instance, to lowest order in 
$u_\pm$ the r.h.s.~of the first equation in~\Ref{susy33} yields 
\ba
u_\pm^2 \hD_\pm \big(u_\pm^{-2} \e^I_{0\pm} \big)
  + \ft12 \,\Gamma^I_{A\dA} \hP_\pm^A \e^\dA_{1\pm} &=& \non
&& \!\!\!\!\!\!\!\!\!\!\!\!\!\!\!\!\!\!\!\!\!\!\!\!\!\!\!\!\!
\!\!\!\!\!\!\!\!\!\!\!\!\!\!\!\!\!\!\!\!\!\!\!\!\!\!\!\!\!
= \; (u_\pm^{-2}\!-\!1)\, \rho^{-1}\partial_\pm\rho \, \e^I_\pm
  + D_\pm \e^I_\pm + \ft12 \big( \rho^{-1} \partial_\pm \rho \big)^{-1} 
   P_\pm^A P_\pm^A\,\e^I_{\pm}
  \non[.5ex]
  && 
\!\!\!\!\!\!\!\!\!\!\!\!\!\!\!\!\!\!\!\!\!\!\!\!\!\!\!\!\!
\!\!\!\!\!\!\!\!\!\!\!\!\!\!\!\!\!\!\!\!\!\!\!\!\!\!\!\!\!
=\;    (u_\pm^{-2}\!-\!1)\, \rho^{-1}\partial_\pm\rho \, \e^I_\pm
  + D_\pm \e^I_\pm + \partial_\pm\sigma\,\e^I_{\pm}
  \non[.5ex]
  &&
\!\!\!\!\!\!\!\!\!\!\!\!\!\!\!\!\!\!\!\!\!\!\!\!\!\!\!\!\!
\!\!\!\!\!\!\!\!\!\!\!\!\!\!\!\!\!\!\!\!\!\!\!\!\!\!\!\!\!
=\;      (u_\pm^{-2}\!-\!1)\, \delta_\e\,\psi_{2\,\pm}^I 
  \mp\delta_\e\,\psi_{\pm}^I 
\;,
\nn
\ea
where for the second equality we have used the bosonic part of the 
conformal constraint~\Ref{TTT}. Likewise, the r.h.s. of the second 
equation in~\Ref{susy33} combines contributions from both terms in 
order $u_\pm^{-1}$. The second condition \Ref{ansatzE} is thus 
crucial for our scheme to work.

When checking \Ref{Dirac1} against the fermionic equations of
motion of $N=16$ supergravity to linear order, it is sufficient
to use the bosonic linear system \Ref{Lbos}. At higher order in
the fermions, the fermionic bilinears in \Ref{LS1} will become 
relevant, and modify the r.h.s. of \Ref{Dirac1} by cubic fermionic 
terms. It is tempting to speculate that these are precisely the higher 
order fermionic terms in the fermionic field equations; however, the
corresponding corrections appear only at constant or higher order 
in the $u_\pm$. 

Further confirmation for the correctness of the ansatz \Ref{fermions} 
comes from the fact that the singular fermionic contributions in the 
full connection of the linear system~\Ref{LS3} all arise from expanding
the fermionic bilinear
\be
\Big(
\Psi_\pm^I (u_\pm) \Psi_\pm^J (u_\pm) 
-\ft14\Gamma^{IJ}_{\dA\dB}\,\CX^\dA(u_\pm)\CX^\dB(u_\pm)\Big)\,X^{IJ}
-2
\Gamma^I_{A\dA} \Psi^I_\pm(u_\pm) \CX^\dA_\pm(u_\pm)\, Y^A
\;,
\ee
in lowest order. This means that the product of two spinorial 
representations contains the adjoint representation of $\KE$.

\section{Outlook}
Much attention has been devoted recently to the 
possible emergence of the indefinite Kac Moody algebras ${\rm E}_{10}$ 
\cite{DaHeNi02,BrGaHe04} and $E_{11}$ \cite{West01,EngHou03}
and their relevance to the bosonic sector of M theory. Although
no similar treatment exists for the fermionic sector, it is natural 
to conjecture that the fermionic degrees of freedom of M theory
should consequently transform as spinors (i.e.~as double-valued 
representations) under the maximal compact subgroups of these
Kac Moody groups, in accordance with the chain of embeddings of 
`generalized R symmetries'
\be
\dots \subset {\rm Spin}(16) \subset \KE \subset K({\rm E}_{10})
\subset  \dots
\ee
We believe our results strengthen the evidence that these groups
are indeed the correct R symmetry groups not only of dimensionally
reduced supergravity, but possibly even of M theory itself. To work
out the relevant spinor representations for $K({\rm E}_{10})$ (and
also for $K({\rm E}_{11})$) will be no easy task; a recursive approach 
based an expansion under the respective subgroups ${\rm Spin}(10)$ and 
${\rm Spin}(1,10)$ is expected to generate infinite towers of similar 
complexity as those in \cite{NicFis03}. Still, whatever these representations 
are, it is clear that the $\KE$ spinor representations studied here 
must be embeddable in these bigger representations.

In recent work~\cite{DufLiu03,Hull03,BDLW03}, an alternative chain of 
finite dimensional `generalized holonomy groups' \footnote{The group
${\rm SL}(32,\R)$ was already suggested as a symmetry in \cite{BaeWes00}.}
\be
\dots \subset {\rm SO}(16) ~\subset~ {\rm SO}(16)_+\times 
{\rm SO}(16)_- ~\subset~ {\rm SO}(32) ~\subset~ {\rm SL}(32,\R)\;,
\ee
was proposed to arise in the reduction of M theory to $d=2,1,0$
dimensions (with analogous chains for spacelike and null reductions).
This embedding chain is suggested by the fact that the $D\!=\!11$
$\Gamma$-matrices generate ${\rm SL}(32,\R)$, and therefore all of the
subgroups listed above. For the case $d=2$, we have found that the
chiral split of $\KE$ with regard to the branch points $\sr = \pm 1$
does seem to suggest a hidden ${\rm Spin}(16)_+\times {\rm
Spin}(16)_-$ symmetry \footnote{Or a hidden ${\rm SO}(16,\C)$ for the
reduction with one timelike Killing vector, as suggested in
\cite{DufLiu03}, and in agreement with the fact that the branch points
are located at $\sr = \pm i$ for a Euclidean worldsheet. Different
real forms of $\KE$ have been recently studied
in~\cite{Keur04}.}. Indeed, the transformation of the supersymmetry
constraint in \Ref{dSI} shows how to realize this group via an
unfaithful and on shell realization of $\KE$.  However, the two chains
no longer match because ${\rm Spin}(16)_+\times {\rm Spin}(16)_-$ is
not a subgroup of $\KE$.

The situation is much less clear for the conjectured holonomy groups
${\rm SO}(32)$ and ${\rm SL}(32,\R)$. Although the 32-component Majorana 
spinor parameter of $D\!=\!11$ supergravity can be assigned 
to the $\bf{32}$ representation of ${\rm SO}(32)$ or ${\rm SL}(32,\R)$, 
no such assignment is possible for the gravitino: neither the 
$\bf{288}$ of ${\rm Spin}(10)$ nor the $\bf{320}$ of ${\rm Spin}(1,10)$ 
can be `lifted' to a representation of ${\rm SO}(32)$ or ${\rm SL}(32,\R)$, 
respectively. Moreover, as shown in \cite{Keur03}, these groups 
do not lead to the required double-valued representations when 
oxidized back to $d>2$ dimensions. This is obvious for ${\rm SL}(32,\R)$,
which does not possess finite dimensional double valued 
representations,  but it is also easy to see that no representation 
of ${\rm SO}(32)$ can ever give rise to a spinorial representation under 
its diagonally embedded ${\rm SO}(16)$ subgroup. Similar comments apply 
concerning the relation of ${\rm SO}(32)$ and ${\rm SL}(32,\R)$ 
to the involutory 
subgroups $K({\rm E}_{10})$ and $K({\rm E}_{11})$; although, at 
levels $\ell \leq 4$ the latter contain all the requisite ${\rm SO}(10)$ 
and ${\rm SO}(1,10)$ representations arising in the decomposition of 
${\rm SO}(32)$ and ${\rm SL}(32,\R)$, respectively, these do not close into 
finite subalgebras of either $K({\rm E}_{10})$ or $K({\rm E}_{11})$.

\subsection*{Acknowledgements} 
This work was begun during the {\em ESI Workshop on gravity in two dimensions}
in September 2003. We thank the Erwin Schr\"{o}dinger Institute in Vienna 
for hospitality and partial support.


\begin{thebibliography}{10}

\bibitem{CreJul79}
E.~Cremmer and B.~Julia, { The ${SO}(8)$ supergravity},  { Nucl. Phys.} { B159}
  (1979)
141--212.

\bibitem{Juli80}
B.~Julia, in {Superspace and Supergravity}, eds. S.W.~Hawking and M.~Rocek
(Cambridge University press, 1980)

\bibitem{Kac}
V.G.~Kac, {Infinite dimensional Lie algebras}, Cambridge University
Press, 3rd edition (1990)

\bibitem{PreSeg86}
A.~Pressley and G.~Segal, { Loop groups}.
\newblock Oxford Mathematical Monographs. The Clarendon Press Oxford University
  Press, New York, 1986.
\newblock Oxford Science Publications.

\bibitem{Berm89}
S.~Berman, { On generators and relations for certain involutory subalgebras of
  {K}ac-{M}oody {L}ie algebras},  { Commun. in Algebra} { 12} (1989)
  3165--3185.

\bibitem{DieHoe84}
C.~Hoenselars and W.~Dietz (eds.): {Solutions of Einstein's Equations:
Techniques and Results}, Springer Verlag (Berlin), 1984)

\bibitem{BreMai87}
P.~Breitenlohner and D.~Maison, { On the {G}eroch group},  { Ann. Inst. H.
  Poincar{\'e}. Phys. Th{\'e}or.} { 46} (1987)
215--246.

\bibitem{Nico87a}
H.~Nicolai, { The integrability of {$N$}=16 supergravity},  { Phys. Lett.} {
  B194} (1987)
402--407.

\bibitem{NicWar89}
H.~Nicolai and N.~P. Warner, { The structure of {$N$}=16 supergravity in two
  dimensions},  { Commun. Math. Phys.} { 125} (1989) 369--384.

\bibitem{NicSam98}
H.~Nicolai and H.~Samtleben, { Integrability and canonical structure of $d =
  2$, ${N}=16$ supergravity},  { Nucl. Phys.} { B533} (1998) 210--242,
[\href{http://xxx.lanl.gov/abs/hep-th/9804152}{{\tt hep-th/9804152}}].

\bibitem{Nico91}
H.~Nicolai, { Two-dimensional gravities and supergravities as integrable
  systems},  in { Recent Aspects of Quantum Fields} (H.~Mitter and
  H.~Gausterer, eds.), (Berlin), Springer-Verlag, 1991.

\bibitem{JulNic96}
B.~Julia and H.~Nicolai, { Conformal internal symmetry of $2$-$d$ sigma models
  coupled to gravity and a dilaton},  { Nucl. Phys.} { B482} (1996) 431--465,
[\href{http://xxx.lanl.gov/abs/hep-th/9608082}{{\tt hep-th/9608082}}].

\bibitem{Juli83}
B.~Julia, { Application of supergravity to gravitation theories},  in { Unified
  field theories in more than 4 dimensions} (V.~D. Sabbata and E.~Schmutzer,
  eds.), (Singapore), pp.~215--236, World Scientific, 1983.

\bibitem{Koro89}
D.~Korotkin, { Finite-gap solutions of the stationary axisymmetric einstein
  equation in vacuum},  { Theor. Math. Phys.} { 77} (1989)
1018--1031.

\bibitem{KorMat90}
D.~Korotkin and V.~Matveev, { Algebro-geometric solutions of the gravitational
  equations},  { Leningrad Math. J.} { 1} (1990)
379--408.

\bibitem{KorNic94}
D.~Korotkin and H.~Nicolai, { The {E}rnst equation on a {R}iemann surface},  {
  Nucl. Phys.} { B429} (1994) 229--254.

\bibitem{BelZak78}
V.~Belinskii and V.~Zakharov, { Integration of the {E}instein equations by
  means of the inverse scattering problem technique and construction of exact
  soliton solutions},  { Sov. Phys. JETP} { 48} (1978)
985--994.

\bibitem{Keur03}
A.~Keurentjes, { The topology of {U}-duality (sub-)groups},
\href{http://xxx.lanl.gov/abs/hep-th/0309106}{{\tt hep-th/0309106}}.

\bibitem{DaHeNi02}
T.~Damour, M.~Henneaux, and H.~Nicolai, { ${E}_{10}$ and a 'small tension
  expansion' of {M} theory},  { Phys. Rev. Lett.} { 89} (2002) 221601,
[\href{http://xxx.lanl.gov/abs/hep-th/0207267}{{\tt hep-th/0207267}}].

\bibitem{BrGaHe04}
J.~Brown, O.~J. Ganor, and C.~Helfgott, { M-theory and ${E}_{10}$: {B}illiards,
  branes, and imaginary roots},
\href{http://xxx.lanl.gov/abs/hep-th/0401053}{{\tt hep-th/0401053}}.

\bibitem{West01}
P.~C. West, { ${E}_{11}$ and {M} theory},  { Class. Quant. Grav.} { 18} (2001)
  4443--4460,
[\href{http://xxx.lanl.gov/abs/hep-th/0104081}{{\tt hep-th/0104081}}].

\bibitem{EngHou03}
F.~Englert and L.~Houart, { ${{\cal G}}^{+++}$ invariant formulation of gravity
  and {M}-theories: {E}xact {BPS} solutions},  { JHEP} { 01} (2004) 002,
[\href{http://xxx.lanl.gov/abs/hep-th/0311255}{{\tt hep-th/0311255}}].

\bibitem{NicFis03}
H.~Nicolai and T.~Fischbacher, { Low level representations for ${E}_{10}$ and
  ${E}_{11}$},  in { Proceedings of the Ramanaujan International Symposium on
  Kac-Moody Lie Algebras and Applications} (N.~Sthanumoorthy and K.~Misra,
  eds.), vol.~343 of { Contemporary Mathematics}, American Mathematical
  Society, 2003.
\newblock
\href{http://xxx.lanl.gov/abs/hep-th/0301017}{{\tt hep-th/0301017}}.
\newblock

\bibitem{DufLiu03}
M.~J. Duff and J.~T. Liu, { Hidden spacetime symmetries and generalized
  holonomy in {M}-theory},  { Nucl. Phys.} { B674} (2003) 217--230,
[\href{http://xxx.lanl.gov/abs/hep-th/0303140}{{\tt hep-th/0303140}}].

\bibitem{Hull03}
C.~Hull, { Holonomy and symmetry in {M}-theory},
\href{http://xxx.lanl.gov/abs/hep-th/0305039}{{\tt hep-th/0305039}}.

\bibitem{BDLW03}
A.~Batrachenko, M.~J. Duff, J.~T. Liu, and W.~Y. Wen, { Generalized holonomy of
  {M}-theory vacua},
\href{http://xxx.lanl.gov/abs/hep-th/0312165}{{\tt hep-th/0312165}}.

\bibitem{BaeWes00}
O.~B\"arwald and P.C.~West, {Brane rotating symmetries and the fivebrane 
equations of motion}, {Phys. Lett.} { B476} (2000) 157--164
[\href{http://xxx.lanl.gov/abs/hep-th/9912226}{{\tt hep-th/9912226}}].

\bibitem{Keur04}
A.~Keurentjes, { Time-like {T}-duality algebra},
\href{http://xxx.lanl.gov/abs/hep-th/0404174}{{\tt hep-th/0404174}}.

\end{thebibliography}

\providecommand{\href}[2]{#2}\begingroup\raggedright\endgroup

\end{document}